\documentclass[10pt,twocolumn,letterpaper]{article}
\usepackage{cvpr}
\usepackage{amsthm,mathtools,bm}
\usepackage{multirow}
\usepackage{microtype}
\usepackage[skins]{tcolorbox}
\usepackage[above,below]{placeins}
\definecolor{promptsysbg}{HTML}{FFF6D5}\definecolor{promptsysfr}{HTML}{E0B84A}\definecolor{promptusrbg}{HTML}{E3F4E1}\definecolor{promptusrfr}{HTML}{6BAF64}
\newtcolorbox{promptbox}[2]{enhanced,colback=prompt#1bg,colframe=prompt#1fr,coltext=black,arc=2pt,boxrule=0.6pt,left=3pt,right=3pt,top=4pt,bottom=2pt,
  before skip=6pt,after skip=0pt,fontupper=\scriptsize,title={#2},fonttitle=\scriptsize\bfseries,coltitle=black,
  attach boxed title to top left={xshift=5pt,yshift=-\tcboxedtitleheight/2},boxed title style={colback=prompt#1bg,colframe=prompt#1fr,boxrule=0.6pt,arc=1.5pt,left=2pt,right=2pt,top=0.5pt,bottom=0.5pt}}
\definecolor{cvprblue}{rgb}{0.21,0.49,0.74}
\usepackage[pagebackref,breaklinks,colorlinks,allcolors=cvprblue]{hyperref}

\def\confName{CVPR}
\def\confYear{2027}

\title{Agentic-Level Adversarial Purification: Revisiting Diffusion Purifiers\\on Native-Resolution Agentic Inputs}

\author{
Yan Lin$^{1}$\thanks{Equal contribution.}\qquad
Zhongliang Guo$^{2}$\textsuperscript{\textasteriskcentered}\thanks{Corresponding author.}\qquad
Jingyu Lin$^{3}$\qquad
Daizong Liu$^{4}$\\[2mm]
$^{1}$Newcastle University\qquad
$^{2}$University of Aberdeen\qquad
$^{3}$Monash University\qquad
$^{4}$Wuhan University
}

\begin{document}
\maketitle

\begin{abstract}
Vision-language agents act on native-resolution renders of multi-element pages, where one perturbed photo or a few altered text strokes can redirect the decision.
Any content provider can craft such a perturbation on the open model behind the agent, so a plug-in purifier is the natural first line of defense.
Yet diffusion purifiers were designed and evaluated at the fixed resolution of the diffusion model on classifier or natural-image benchmarks, and defense evaluations for agents test JPEG, blur and noise but no diffusion purifier.
We present the first study of adversarial purification at the agentic level, for an agent built on Qwen3-VL-8B: purifiers run at native resolution on multi-element inputs, face grey-box attackers who know the agent's model but not the defense, and are compared at matched clean cost.
We show that treating agentic inputs as classifier inputs misleads both evaluation and design.
Resizing the input to the resolution of the diffusion model, as existing purifiers do, raises the TextVQA clean cost of the Divide-and-Conquer purifier (DC) from 4.0 to 35.7 points.
On web-shop pages, DC keeps 85.0\% grey-box robust accuracy against web-hardened perturbations, where cheap input transformations at matched clean cost keep at most 40.0\%,
and this robustness comes from the global noise level: input-estimated masks and noise levels add none.
These findings hold on a disjoint held-out set and on a second VLM and yield three requirements for defenses; we release the protocol.
\end{abstract}

\section{Introduction}
\label{sec:intro}

Vision-language models (VLMs) now drive agents that read a screenshot or a photograph at its native resolution and act on it, for example by choosing a product on a rendered web page~\cite{koh2024visualwebarena,bai2025qwen3}.
Third parties supply much of this content, and an adversarial perturbation of one image on the page can change the action~\cite{wu2025dissecting,seip2026preference,qian2026penny,han2026adversarial}.
Any content provider can craft such a perturbation on the open model behind the agent, so a plug-in purifier that cleans the input before the agent reads it is the natural first line of defense~\cite{nie2022diffusion}.

\begin{figure}[t]
  \centering
  \includegraphics[width=\linewidth]{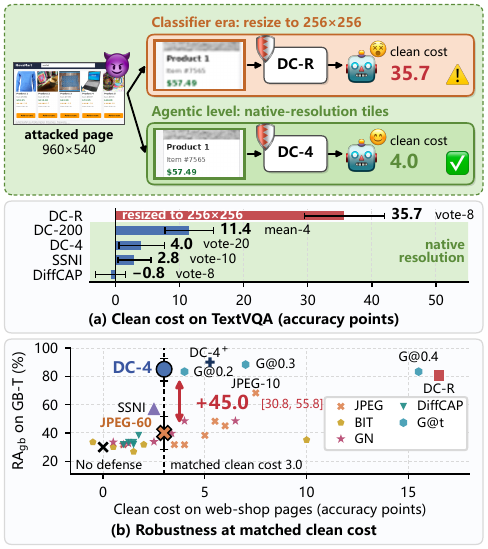}
  \caption{\textbf{At native resolution, a diffusion purifier keeps the text an agent reads and the product it picks}.
  Top: TextVQA clean costs.
  (b)~Perceptible tier. Intervals: 95\%.}
  \label{fig:teaser}
\end{figure}

In this paper, we study adversarial purification at the agentic level: a plug-in pixel-space purifier preprocesses the native-resolution input of a VLM agent.
We call agentic inputs the native-resolution, multi-element inputs on which a VLM agent takes a single-step decision, here a rendered web-shop page with five product cards and a text-heavy photograph from TextVQA~\cite{singh2019towards}.
In the main threat model, a grey-box (purifier-unaware) attacker knows the agent's VLM and task but not the purifier, and perturbs the input within an imperceptible or a perceptible $\ell_\infty$ budget.
We also evaluate same-page adaptive attacks, optimized through the defense.

Existing pixel-space defenses for VLM inputs can be divided into three categories: fixed input transformations such as JPEG, blur, noise~\cite{seip2026preference,qian2026penny,han2026adversarial} and bit-depth reduction~\cite{guo2018countering}, which we call cheap transformations, noise-level-calibrated diffusion purifiers inherited from classifiers~\cite{nie2022diffusion,pei2025divide,sun2025sample}, and VLM-specific diffusion purifiers such as DiffCAP~\cite{fu2026diffusion}.
In the second category, DiffPure sets one global noise level, while Divide-and-Conquer (DC) and SSNI estimate a mask or a level from the input.
However, the clean cost (the accuracy lost on clean inputs) and the robustness of these defenses in front of an agent are judged from evidence gathered under other conditions.
Existing diffusion purifiers run at the fixed resolution of the diffusion model~\cite{pei2025divide,wang2025safeguarding}, and evaluations on agents test input transformations but no diffusion purifier~\cite{seip2026preference,qian2026penny,han2026adversarial}.
The clean cost that a diffusion purifier imposes on an agent, and the robustness it adds over these transformations on the same inputs, have not been measured.
We argue that this defect stems from one latent assumption, which we call the classifier-input assumption: the input is a fixed-resolution image whose evidence is spread over the whole image, and the attack is a small perturbation that the preprocessing in front of the model does not shape.
It does not hold for agentic inputs, where the decision rests on one decisive element, such as a product photo or a few text strokes, and where deployable perturbations are perceptible and shaped by the preprocessing they must pass.
As shown in \cref{fig:teaser}, a diffusion purifier at native resolution keeps both the text an agent reads and the product it picks.
On TextVQA, resizing the input to the resolution of the diffusion model, as DC's official pipeline does (DC-R), costs 35.7\,{\footnotesize[29.5,\,42.0]} percentage points of accuracy, and at native resolution the cost depends on the readout, the rule that turns several stochastic draws into one decision.
Against perceptible perturbations hardened to survive web preprocessing (web-hardened) on web-shop pages, DC at its official noise levels with four function evaluations (DC-4) keeps a grey-box robust accuracy of 85.0\,{\footnotesize[76.7,\,92.5]}.
The highest of the cheap transformations at matched clean cost (settings that cost no more than \mbox{DC-4}) keeps 40.0\,{\footnotesize[28.3,\,51.7]}, a gap of 45.0\,{\footnotesize[30.8,\,55.8]} points.
The resize cost and this gap hold on a disjoint held-out set and on a second VLM, Qwen2.5-VL-7B~\cite{bai2025qwen25vl}.
In \cref{fig:teaser}(b), the most robust setting of the cheap transformations, JPEG-10, keeps 68.3 at a clean cost of 7.5 points, while DC-4 keeps 85.0 at 3.0.

Because the classifier-input assumption does not hold on agentic inputs, purification at the agentic level has to meet three requirements (\cref{sec:requirements}).
The purifier should (i) run at native resolution, with a clean cost that is measured on the decisive element at the readout the agent uses.
It should (ii) neutralize web-hardened perturbations on that element without estimating their strength or location, since such estimates add no robustness over a global noise level.
It should (iii) hold against attackers that know the defense and beat cheap transformations at matched clean cost.

\begin{figure*}[t]
  \centering
  \includegraphics[width=\linewidth]{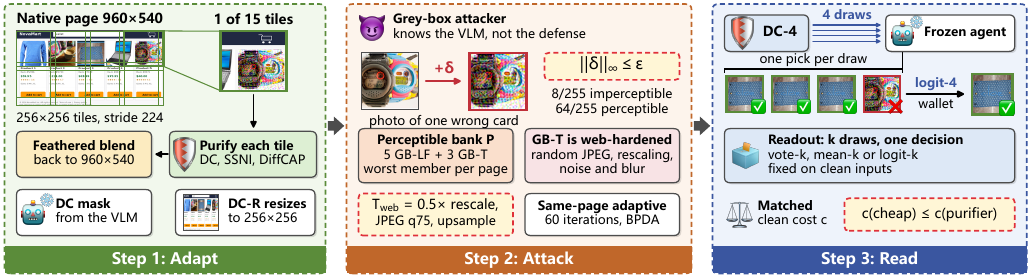}
  \caption{\textbf{AgenticPure fixes the input resolution, the attack set and the readout}.}
  \label{fig:protocol}
\end{figure*}

As illustrated in \cref{fig:protocol}, we introduce AgenticPure, an evaluation protocol with three steps.
Adapt runs DC, SSNI and DiffCAP at native resolution, Attack builds grey-box banks of imperceptible and web-hardened perceptible perturbations and a same-page adaptive attack, and Read compares defenses at matched clean cost, each under a readout fixed on clean inputs.

In summary, our main contributions are as follows:
\begin{enumerate}
  \item We present the first study of adversarial purification at the agentic level and show that the classifier-input assumption misleads both evaluation and design.
  Resolution handling and the readout set the clean cost on text, and DC beats cheap transformations on web-hardened perturbations.
  \item We introduce AgenticPure, an evaluation protocol for agentic-level purification with native-resolution versions of DC, SSNI and DiffCAP, web-hardened grey-box banks, a same-page adaptive attack, comparisons at matched clean cost and held-out confirmation.
  \item We trace where purification succeeds and fails on both agentic inputs: the decision, the attack surface and the clean cost fall on one element,
  and, on web-shop pages, input-estimated masks and noise levels add no robustness over a fixed global noise level at matched clean cost,
  which yields three requirements.
\end{enumerate}

\section{Related Work}
\label{sec:related}

\noindent\textbf{Attacks on VLMs and VLM agents.}
Adversarial images attack VLMs in white-box and transfer settings~\cite{zhao2023evaluating,schlarmann2023adversarial,qi2024visual,bailey2024image}.
Agents on rendered pages are attacked with norm-bounded perturbations~\cite{wu2025dissecting,seip2026preference,qian2026penny,han2026adversarial}, image patches~\cite{aichberger2025mip,liu2026agenthijack} and injected page content~\cite{liao2025eia,ma2025caution,zhang2025attacking}.
The norm-bounded evaluations use $\ell_\infty$ budgets of 8/255 to 16/255 and attacks that are, with one exception~\cite{seip2026preference}, not optimized against the tested defense.
Their pixel-level defenses are input transformations (JPEG, blur, noise)~\cite{seip2026preference,qian2026penny,han2026adversarial}, and PRAC~\cite{seip2026preference} notes that diffusion purification is untested for GUI environments.

\noindent\textbf{Diffusion purification and its evaluation.}
DiffPure~\cite{nie2022diffusion} noises the input to a fixed level and denoises it with a pretrained diffusion model; later purifiers add guidance~\cite{wang2022guided}, vote over several draws~\cite{xiao2023densepure} or purify in one step~\cite{lei2025instant}.
Recent purifiers adapt the noise to the input: DC~\cite{pei2025divide} applies a higher level inside an attention mask, SSNI~\cite{sun2025sample} estimates one level per sample, and MANI-Pure~\cite{huang2025mani} injects frequency-targeted noise.
Adaptive and transfer attacks re-examined purifier robustness~\cite{lee2023robust,kang2023diffattack,wang2024diffhammer,wang2026shared}, and Liu~\etal~\cite{liu2025towards} argue that stochasticity is its primary source.
DiffBreak~\cite{kassis2025diffbreak} evaluates purification under a majority vote over several purified copies.
DC and SSNI were designed and evaluated against norm-bounded imperceptible perturbations ($\ell_\infty$ up to 8/255, $\ell_2$ 0.5); we evaluate them on agentic inputs, against imperceptible and perceptible perturbations.

\noindent\textbf{Purification for VLMs.}
DiffCAP~\cite{fu2026diffusion} sets the noise level with a stopping rule on image embeddings, and DiffPure-VLM~\cite{wang2025safeguarding} resizes the input to $256 \times 256$.
F3~\cite{zhang2025fighting} is a noise-based purifier independent of the input resolution, and PureProof~\cite{cao2026pureproof} attacks VLMs, including \mbox{Qwen3-VL-8B}, behind DiffPure and GDMP~\cite{wang2022guided} on ImageNet images.
\mbox{DiffCAP} reports that DiffPure lowers the TextVQA accuracy of \mbox{LLaVA-1.5-7B} from 37.1 to 20.9, but none of these works reports agent screenshots at native resolution, where we measure the clean cost.

\noindent\textbf{Input transformations.}
JPEG compression~\cite{dziugaite2016study,das2017keeping} and bit-depth reduction~\cite{xu2018feature,guo2018countering} were proposed for classifiers, and adaptive attacks circumvent defenses of this kind on classifiers~\cite{athalye2018obfuscated,tramer2020adaptive} and on MiniGPT-4~\cite{zhang2025self}.
In PRAC~\cite{seip2026preference}, JPEG at quality 75 lowers the attack success rate on Qwen3-VL from 93.3\% to 21.6\%, and optimizing through JPEG restores it to 53.3\%.
AgenticPure compares these transformations with diffusion purifiers at matched clean cost, against perturbations optimized to survive web preprocessing.

\section{Agentic-Level Setting and the AgenticPure Protocol}
\label{sec:setting}

\subsection{Agentic Inputs and Threat Model}
\label{sec:inputs}

An input is \emph{agentic} when it is a native-resolution, multi-element input on which a VLM agent takes a single-step decision.
In TextVQA~\cite{singh2019towards}, the agent answers a question about a photograph at its native resolution.
In the web shop, it is asked for a product and chooses one of five product cards on a rendered $960\times540$ page, as in PRAC~\cite{seip2026preference}.
The \emph{decisive element} is the region whose content fixes the decision: the scene text or the product photos.
Unless marked, results are for an agent built on Qwen3-VL-8B~\cite{bai2025qwen3}, which we call the VLM.
\Cref{supp:vlm2} evaluates both tasks on a second VLM, Qwen2.5-VL-7B~\cite{bai2025qwen25vl}.

The attacker adds a perturbation of $\ell_\infty$ norm at most $\epsilon$, at the \emph{imperceptible tier} ($\epsilon = 8/255$) or the \emph{perceptible tier} ($\epsilon = 64/255$).
It covers the TextVQA image or, on a web-shop page, the photo of one wrong card, the attacker's target.
A \emph{grey-box} attacker knows the VLM and the task but not the defense.
A \emph{same-page adaptive} attacker also knows the defense and optimizes through it on the evaluated page.
The \emph{web pipeline} $T_{\mathrm{web}}$ downsamples the perturbed photo by $0.5\times$, compresses it with JPEG at quality 75, and upsamples it back.
An attack succeeds when the agent picks the target card or, on TextVQA, when answer accuracy drops.

\subsection{Purifiers and Cheap Transformations}
\label{sec:purifiers}

Every \emph{diffusion purifier} follows one pipeline: resize the input to $256\times256$ or tile it, add noise up to a \emph{noise level} $t\in[0,1]$ (a fraction of the diffusion horizon), denoise with a diffusion model~\cite{dhariwal2021diffusion}, and pass the result to the VLM.

Divide-and-Conquer (DC)~\cite{pei2025divide} noises an attention-masked region to $t_{l} = 0.4$ and the rest to $t_{s} = 0.2$, its official ImageNet levels.
DC-200 is DC with its official 200-step denoising loop~\cite{pei2025divide}, tiled at native resolution.
Its mask is the top 15\% of DC's attention map, computed on the VLM's vision-tower activations, and it runs in mixed precision with four draws instead of a 20-member ensemble.
DC-4 is DC at its official noise levels with four function evaluations, tiled at native resolution (its timestep grid is given in \cref{supp:impl}).
\mbox{DC-R} runs the loop of DC-4 on the input resized to $256\times256$, as in DC's official pipeline.
DC-4\textsuperscript{+} is DC-4 with a stronger level (0.30) in the unmasked region.
SSNI~\cite{sun2025sample} runs at its published hyperparameters and DiffCAP~\cite{fu2026diffusion} follows its Alg.~1 and~2 on CLIP features of the tiles, both with a two-step DDIM~\cite{song2021denoising} reverse process.
G@$t$ is diffusion purification at a fixed global noise level $t$ (two DDIM evaluations).
On the data and classifiers of the original papers and without tiling, DC scores 69.5 against a published 74.5 and SSNI 72.3 against 70.3 on ImageNet~\cite{deng2009imagenet} ($n=256$).
DiffCAP at its published setting scores 83.1 and 80.5 on Caltech-101~\cite{fei2004learning} and 86.6 and 80.4 on \mbox{ImageNet-R}~\cite{hendrycks2021many}, clean and under attack, against published values of 82.6, 80.9, 87.2 and 81.1 ($n=1000$).
Every published accuracy lies inside the 95\% Wilson interval of ours (\cref{supp:tab:portcheck}).
Our implementations reproduce the published purifiers, so the effects reported below come from the agentic setting.

The \emph{cheap transformations} are three fixed input transformations: JPEG compression at quality $q$ (JPEG-$q$)~\cite{das2017keeping}, bit-depth reduction to $b$ bits (BIT-$b$)~\cite{guo2018countering}, and additive Gaussian noise of standard deviation $s/255$ (GN-$s$).
Their grids span $q\in[10,95]$, $b\in[2,7]$ and $s\in[1,32]$.
No defense is the undefended agent.

\subsection{The AgenticPure Protocol}
\label{sec:protocol}

As shown in \cref{fig:protocol}, AgenticPure has three steps, which fix the input resolution, the attack set, and the readout.

\noindent\textbf{Adapt.}
Every purifier except DC-R runs on overlapping $256\times256$ tiles of the native-resolution input (stride 224, 32-pixel feathered blend), and DC takes its attention mask from the VLM.

\noindent\textbf{Attack.}
The perceptible bank $\mathcal{P}$ holds five low-frequency (GB-LF) and three \emph{web-hardened} members (\mbox{GB-T}), which are optimized to survive random JPEG compression, rescaling, noise and blur~\cite{athalye2018synthesizing}.
The imperceptible TextVQA bank holds PGD, APGD~\cite{croce2020reliable}, two GB-LF members, GB-T and a member optimized through JPEG at quality 75, and \mbox{GB-T10} adds a member hardened on JPEG qualities down to 10.
All members are generated on the undefended VLM, and each defense is scored, per page, on its worst member.
The \emph{web-surviving subset} holds the pages where a member of $\mathcal{P}$ still hits its target after $T_{\mathrm{web}}$.
The same-page adaptive attack starts from the grey-box member and runs 60 iterations through the defense, with BPDA~\cite{athalye2018obfuscated} for non-differentiable steps.
A convergence check continues it to 180 iterations, and a gradient-masking check tests that accuracy does not rise with the budget.

\noindent\textbf{Read.}
A \emph{draw} is one stochastic purification of one input, and a \emph{readout} maps the draws to one score.
Following the majority vote of DiffBreak~\cite{kassis2025diffbreak}, vote-$k$ takes the plurality decision over $k$ draws, mean-$k$ the mean score of the $k$ draws, and logit-$k$ the decision of the mean logit.
Each defense uses, per task, the readout with the highest clean accuracy on the development inputs.
\emph{At matched clean cost}, a cheap transformation or fixed noise level is the grid setting with the largest clean cost not above the purifier's.
Every matched setting is thus at least as accurate on the clean development inputs as the purifier it is compared with.

\subsection{Metrics and Statistics}
\label{sec:metrics}

\emph{Clean cost} is the accuracy of No defense minus the accuracy under the defense on clean inputs, in points.
RA\textsubscript{gb} is the accuracy under the grey-box bank, and RA\textsubscript{ad}, the accuracy under the same-page adaptive attack, is an upper bound on robust accuracy.
Intervals are 95\% paired bootstrap intervals~\cite{efron1979bootstrap} (10,000 resamples), written $v$\,{\footnotesize[$l$,\,$u$]}.
Attack success rates are proportions with 95\% Wilson intervals~\cite{wilson1927probable}.
The development set has 250 TextVQA questions, 200 clean web-shop pages and 60 attacked pages.
The disjoint held-out set has 1,000 questions (with a fixed subset of 250) and 300 pages.
Held-out attacks use the pages that No defense answers correctly.
Results are on the development set unless marked.
Two annotators assign perturbed product photos to their class at least as accurately as clean ones (200 perturbed and 100 clean photos, class agreement $\kappa = 0.90$~\cite{cohen1960coefficient}).
Perceptible perturbations leave the products recognizable, so the correct choice stays well defined.
On the held-out set, tests with Holm's correction~\cite{holm1979simple} at level 0.05 confirm that DC-R costs more than 20 points and SSNI and \mbox{DiffCAP} less than 8 points on TextVQA.
On web-shop pages, they confirm that DC-200 keeps more RA\textsubscript{gb} than the matched cheap transformations, and that the RA\textsubscript{gb} of DC-4 with and without its mask differs by less than 5 points.

\section{Revisiting Diffusion Purifiers at the Agentic Level}
\label{sec:revisit}
Classifier-era evaluations of diffusion purifiers~\cite{nie2022diffusion,lee2023robust} share one pipeline: the input is brought to the resolution of the diffusion model, noised, denoised, and passed to the model.
It hides the classifier-input assumption, which we test in parts: a fixed resolution and the readout (\cref{sec:res}), a small perturbation that preprocessing does not shape (\cref{sec:tier}), an estimable strength and location (\cref{sec:components}), and evidence spread over the image (\cref{sec:element}).
\Cref{sec:adaptive} tests what an attacker who knows the defense achieves.
\subsection{Impact of Resolution Handling and the Readout}\label{sec:res}
Existing diffusion purifiers run at the fixed resolution of the diffusion model, and DC~\cite{pei2025divide} and DiffPure-VLM~\cite{wang2025safeguarding} resize the input to $256 \times 256$.
This practice relies on the latent assumption that the content the model needs survives the resize, and reported clean accuracies assume that the way the stochastic draws are read out does not matter.
Neither assumption holds on agentic inputs.

\noindent\textbf{Settings.}
We measure the clean cost of five defenses on TextVQA ($n=250$) and on web-shop pages ($n=200$).
DC-R applies DC-4 to the input resized as in DC's official pipeline, while DC-200, DC-4, SSNI and DiffCAP run on tiles at native resolution.
Each defense is scored at its fixed readout, and DC-4 is additionally scored at mean-4, the TextVQA readout of DC-200.

\begin{table}[t]
  \centering\footnotesize
  \begin{tabular}{@{}l@{\hskip 4pt}l@{\hskip 5pt}r@{\,}l@{\hskip 5pt}r@{\,}l@{\hskip 5pt}r@{}}
\toprule
Defense & Readout & \multicolumn{2}{c}{Development} & \multicolumn{2}{c}{Held-out} & \multicolumn{1}{c}{$n$} \\
\midrule
\multicolumn{7}{@{}l}{\emph{TextVQA (development $n=250$, held-out $n$ at right)}} \\
DC-R & vote-8 & 35.7 & {\scriptsize[29.5,\,42.0]} & 41.5 & {\scriptsize[38.3,\,44.6]} & 1000 \\
DC-200 & mean-4 & 11.4 & {\scriptsize[7.7,\,15.3]} & 11.4 & {\scriptsize[7.8,\,15.1]} & 250 \\
DC-4 & vote-20 & 4.0 & {\scriptsize[0.5,\,7.6]} & 5.9 & {\scriptsize[4.0,\,7.9]} & 1000 \\
SSNI & vote-10 & 2.8 & {\scriptsize[0.4,\,5.5]} & 2.1 & {\scriptsize[$-$0.5,\,4.8]} & 250 \\
DiffCAP & vote-8 & $-$0.8 & {\scriptsize[$-$3.1,\,1.5]} & $-$0.1 & {\scriptsize[$-$2.0,\,1.7]} & 250 \\
\addlinespace[2pt]
DC-4 & mean-4 & 10.8 & \multicolumn{4}{@{}l@{}}{{\scriptsize[6.9,\,14.9]}} \\
\hspace{0.5em}DC-200 $-$ DC-4 & mean-4 & 0.6 & \multicolumn{4}{@{}l@{}}{{\scriptsize[$-$1.5,\,2.6]}} \\
\midrule
Defense & Readout & \multicolumn{2}{c}{Development} & \multicolumn{2}{c}{Held-out} & \multicolumn{1}{c}{Time} \\
\midrule
\multicolumn{7}{@{}l}{\emph{Web shop (development $n=200$, held-out $n=200$ of 300)}} \\
DC-R & logit-4 & 16.5 & {\scriptsize[11.3,\,22.0]} & 17.3 & {\scriptsize[11.0,\,23.8]} & 0.29 \\
DC-200 & logit-4 & 3.0 & {\scriptsize[$-$0.5,\,6.5]} & 0.5 & {\scriptsize[$-$3.0,\,4.5]} & 45.57 \\
DC-4 & logit-4 & 3.0 & {\scriptsize[$-$0.3,\,6.5]} & 1.0 & {\scriptsize[$-$2.5,\,4.5]} & 1.05 \\
SSNI & vote-10 & 2.5 & {\scriptsize[0.0,\,5.5]} & 0.0 & {\scriptsize[$-$3.0,\,3.0]} & 12.35 \\
DiffCAP & vote-8 & 1.0 & {\scriptsize[$-$1.0,\,3.0]} & 0.0 & {\scriptsize[$-$2.5,\,2.5]} & 0.57 \\
\bottomrule
\end{tabular}
  \caption{\textbf{Resolution handling and the readout set the clean cost on text}.
  Points, 95\% intervals; time: seconds per draw.}
  \label{tab:clean}
\end{table}

\noindent\textbf{Results.}
As shown in \cref{tab:clean} and \cref{fig:teaser}(a), the clean cost on text-heavy inputs depends on how the purifier handles resolution.
The resize of DC-R costs 35.7\,{\footnotesize[29.5,\,42.0]} points on TextVQA at vote-8.
On Qwen2.5-VL-7B, it costs 58.8\,{\footnotesize[52.5,\,64.7]} points, 32.3\,{\footnotesize[26.3,\,38.4]} more than DC-4.
At native resolution, DC-200 costs 11.4\,{\footnotesize[7.7,\,15.3]} points at mean-4.
On web-shop pages, DC-R costs 16.5\,{\footnotesize[11.3,\,22.0]} points, while DC-200 and DC-4 cost 3.0\,{\footnotesize[$-$0.5,\,6.5]} and 3.0\,{\footnotesize[$-$0.3,\,6.5]} points, both at the logit-4 readout.

The readout changes the clean cost of one and the same purifier.
At mean-4, DC-4 costs 10.8\,{\footnotesize[6.9,\,14.9]} points on TextVQA, and the paired difference of DC-200 minus DC-4 is 0.6\,{\footnotesize[$-$1.5,\,2.6]} points, so both loops cost about 11 points at mean-4.
At its fixed readout of vote-20, DC-4 costs 4.0\,{\footnotesize[0.5,\,7.6]} points.
The readout is thus part of the defense that an evaluation has to fix.
SSNI at vote-10 and DiffCAP at vote-8 cost 2.8\,{\footnotesize[0.4,\,5.5]} and $-$0.8\,{\footnotesize[$-$3.1,\,1.5]} points.
The two DC loops differ in time: one draw on a TextVQA image takes 63.70\,s with DC-200 and 1.41\,s with DC-4.
DC-4 thus matches the clean cost of the 200-step loop at a small fraction of its time.
On held-out TextVQA, DC-R costs 41.5\,{\footnotesize[38.3,\,44.6]} points, DC-200 11.4\,{\footnotesize[7.8,\,15.1]} points at \mbox{mean-4}, and DC-4 only 5.9\,{\footnotesize[4.0,\,7.9]} points at vote-20.
\emph{The clean cost of diffusion purification on the text-heavy inputs of an agent depends on resolution handling and on the readout.}

The cause is that the evidence on these inputs is small text and small elements, which a resize does not preserve, and that the draws of a stochastic purifier disagree on a part of the inputs.

\subsection{Impact of the Perturbation Tier}\label{sec:tier}
Evaluations of pixel-level defenses for agents~\cite{seip2026preference,qian2026penny,han2026adversarial} use $\ell_\infty$ budgets of 8/255 to 16/255, and DC and SSNI are designed for norm-bounded imperceptible perturbations~\cite{pei2025divide,sun2025sample}.
Evaluating at this tier relies on the latent assumption that the perturbation is small and is not shaped by the preprocessing it has to pass through.
On agentic inputs, this assumption does not hold.

\noindent\textbf{Settings.}
At the imperceptible tier ($\epsilon = 8/255$), we attack 126 development TextVQA images that No defense answers with accuracy at least 0.5, and take the worst case over a bank of grey-box perturbations in which GB-T is hardened on JPEG qualities 60 to 95.
GB-T10 adds one member that is hardened on qualities 10 to 95.
The defense is JPEG at qualities 10, 20 and 30, and a differentiable JPEG at quality 20, on which GB-T10 is evaluated.
At the perceptible tier, we attack web-shop pages ($n=60$) and compare DC-4 with JPEG-60, BIT-3 and GN-8, the cheap transformations matched to its clean cost of 3.0 points.
With No defense, the attack success rate is 0.617\,{\footnotesize[0.490,\,0.729]} for GB-T (37 of 60) and 0.517\,{\footnotesize[0.393,\,0.638]} after the web pipeline (31 of 60).

\begin{figure}[t]
  \centering
  \includegraphics[width=\linewidth]{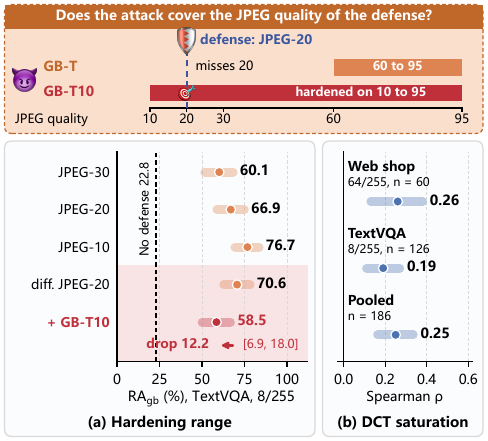}
  \caption{\textbf{JPEG's robust accuracy depends on the attack's hardening range}.
  (a)~diff.: differentiable. (b)~Correlation of predicted DCT saturation with RA\textsubscript{gb}. Intervals: 95\%.}
  \label{fig:tier8}
\end{figure}

\noindent\textbf{Results.}
As shown in \cref{fig:tier8}(a), the robust accuracy of JPEG at the imperceptible tier depends on the range of qualities on which the attack is hardened.
Against the bank with GB-T, JPEG-20 keeps a RA\textsubscript{gb} of 66.9\,{\footnotesize[58.7,\,74.9]} on TextVQA.
JPEG-30 and JPEG-10 keep 60.1 and 76.7: the further the quality lies below the hardening range of \mbox{GB-T}, the higher the RA\textsubscript{gb}.
Adding GB-T10, whose hardening range covers quality 20, lowers the RA\textsubscript{gb} of the differentiable \mbox{JPEG-20} from 70.6\,{\footnotesize[62.7,\,78.3]} to 58.5\,{\footnotesize[50.0,\,66.7]}, a paired drop of 12.2\,{\footnotesize[6.9,\,18.0]} points.
The top of \cref{fig:tier8} shows both hardening ranges: 60 to 95 for GB-T, which misses the defense quality 20, and 10 to 95 for GB-T10.
Part of the robustness of JPEG comes from quantization undoing the perturbation.
In \cref{fig:tier8}(b), the fraction of DCT coefficients that a fixed quantizer is predicted to map back to the quantized clean input correlates with the measured robust accuracy (Spearman correlation 0.19\,{\footnotesize[0.11,\,0.27]} on TextVQA and 0.26\,{\footnotesize[0.13,\,0.38]} on web-shop pages).

\begin{table*}[t]
  \centering\small
  \begin{tabular}{@{}l@{\hskip 12pt}r@{\,}l@{\hskip 12pt}r@{\,}l@{\hskip 12pt}r@{\,}l@{\hskip 12pt}r@{\,}l@{\hskip 12pt}r@{\,}l@{\hskip 12pt}r@{\,}l@{}}
\toprule
 & \multicolumn{8}{c}{Development split} & \multicolumn{4}{c}{Held-out split} \\
\cmidrule(lr){2-9}\cmidrule(l){10-13}
 & \multicolumn{2}{c}{Clean cost} & \multicolumn{2}{c}{Worst of $\mathcal{P}$} & \multicolumn{2}{c}{GB-T} & \multicolumn{2}{c}{$T_{\mathrm{web}}(x_{\mathrm{adv}})$} & \multicolumn{2}{c}{Clean cost} & \multicolumn{2}{c}{Worst of $\mathcal{P}$} \\
Defense & \multicolumn{2}{c}{$n=200$} & \multicolumn{2}{c}{$n=60$} & \multicolumn{2}{c}{$n=60$} & \multicolumn{2}{c}{$n=60$} & \multicolumn{2}{c}{$n=200$} & \multicolumn{2}{c}{$n=273$} \\
\midrule
JPEG-60 & 3.0 & {\scriptsize[1.0,\,5.5]} & 40.0 & {\scriptsize[28.3,\,51.7]} & 40.0 & {\scriptsize[28.3,\,51.7]} & 68.3 & {\scriptsize[56.7,\,80.0]} & 1.0 & {\scriptsize[$-$2.5,\,4.5]} & 44.0 & {\scriptsize[38.1,\,49.8]} \\
BIT-3 & 3.0 & {\scriptsize[0.5,\,6.0]} & 36.7 & {\scriptsize[25.0,\,48.3]} & 40.0 & {\scriptsize[28.3,\,51.7]} & 66.7 & {\scriptsize[55.0,\,78.3]} & 1.5 & {\scriptsize[$-$1.5,\,4.5]} & 39.2 & {\scriptsize[33.3,\,45.1]} \\
GN-8 & 2.5 & {\scriptsize[0.4,\,4.9]} & 30.0 & {\scriptsize[19.2,\,40.8]} & 33.8 & {\scriptsize[22.9,\,45.0]} & 65.8 & {\scriptsize[54.6,\,76.7]} & $-$1.1 & {\scriptsize[$-$4.0,\,1.6]} & 42.7 & {\scriptsize[37.3,\,48.2]} \\
\midrule
DC-200 & 3.0 & {\scriptsize[$-$0.5,\,6.5]} & 88.3 & {\scriptsize[80.0,\,95.0]} & 86.7 & {\scriptsize[76.7,\,95.0]} & 90.0 & {\scriptsize[81.7,\,96.7]} & 0.5 & {\scriptsize[$-$3.0,\,4.5]} & 85.0 & {\scriptsize[80.6,\,89.0]} \\
DC-4 & 3.0 & {\scriptsize[$-$0.3,\,6.5]} & 87.5 & {\scriptsize[79.2,\,94.2]} & 85.0 & {\scriptsize[76.7,\,92.5]} & 90.8 & {\scriptsize[83.3,\,97.5]} & 1.0 & {\scriptsize[$-$2.5,\,4.5]} & 83.2 & {\scriptsize[78.8,\,87.4]} \\
DC-4\textsuperscript{+} & 5.3 & {\scriptsize[1.5,\,9.3]} & 89.2 & {\scriptsize[81.7,\,95.8]} & 90.0 & {\scriptsize[82.5,\,96.7]} & 90.8 & {\scriptsize[83.3,\,96.7]} & 5.3 & {\scriptsize[1.0,\,9.8]} & 88.3 & {\scriptsize[84.8,\,91.8]} \\
DC-R & 16.5 & {\scriptsize[11.3,\,22.0]} & 79.2 & {\scriptsize[69.2,\,88.3]} & 80.8 & {\scriptsize[70.8,\,89.2]} & 80.0 & {\scriptsize[70.0,\,89.2]} & 17.3 & {\scriptsize[11.0,\,23.8]} & 76.6 & {\scriptsize[72.0,\,81.0]} \\
G@0.2 & 4.0 & {\scriptsize[1.0,\,7.5]} & 85.0 & {\scriptsize[75.0,\,93.3]} & 83.3 & {\scriptsize[73.3,\,91.7]} & 91.7 & {\scriptsize[85.0,\,98.3]} & 2.0 & {\scriptsize[$-$1.5,\,5.5]} & 79.5 & {\scriptsize[74.4,\,84.2]} \\
SSNI & 2.5 & {\scriptsize[0.0,\,5.5]} & 56.7 & {\scriptsize[43.3,\,68.3]} & 56.7 & {\scriptsize[43.3,\,68.3]} & 88.3 & {\scriptsize[80.0,\,95.0]} & 0.0 & {\scriptsize[$-$3.0,\,3.0]} & 57.1 & {\scriptsize[51.3,\,63.0]} \\
DiffCAP & 1.0 & {\scriptsize[$-$1.0,\,3.0]} & 30.0 & {\scriptsize[18.3,\,41.7]} & 31.7 & {\scriptsize[20.0,\,43.3]} & 60.0 & {\scriptsize[46.7,\,71.7]} & 0.0 & {\scriptsize[$-$2.5,\,2.5]} & 39.2 & {\scriptsize[33.3,\,45.1]} \\
\bottomrule
\end{tabular}
  \caption{\textbf{On web-shop pages, DC-4 keeps far more RA\textsubscript{gb} than the matched cheap transformations} (top rows).
  Points, 95\% intervals.}
  \label{tab:web}
\end{table*}

As shown in \cref{tab:web} and \cref{fig:teaser}(b), \mbox{DC-4} beats the cheap transformations on web-hardened perturbations at the perceptible tier.
Against GB-T, \mbox{DC-4} keeps a RA\textsubscript{gb} of 85.0\,{\footnotesize[76.7,\,92.5]}, while \mbox{JPEG-60}, BIT-3 and GN-8 keep at most 40.0\,{\footnotesize[28.3,\,51.7]}, a paired gap of 45.0\,{\footnotesize[30.8,\,55.8]} points over the highest of them.
The web pipeline weakens the attack, yet \mbox{DC-4} stays far above the matched cheap transformations.
After the web pipeline itself, \mbox{DC-4} keeps 90.8\,{\footnotesize[83.3,\,97.5]} and the matched cheap transformations keep at most 68.3\,{\footnotesize[56.7,\,80.0]}.
On the 31 pages of the web-surviving subset, \mbox{DC-4} keeps 83.9\,{\footnotesize[70.0,\,96.3]} and the matched cheap transformations at most 45.2\,{\footnotesize[27.6,\,63.0]}.
Scored on the members selected against \mbox{DC-4}, \mbox{DC-200} keeps 86.7\,{\footnotesize[76.7,\,95.0]} against \mbox{GB-T} and 90.0\,{\footnotesize[81.7,\,96.7]} after the web pipeline, on par with \mbox{DC-4}.
On held-out pages, \mbox{DC-4} keeps 83.2\,{\footnotesize[78.8,\,87.4]} against the worst of $\mathcal{P}$ and the matched cheap transformations at most 44.0\,{\footnotesize[38.1,\,49.8]}.
On these pages, \mbox{DC-200} keeps 41.0\,{\footnotesize[35.0,\,46.5]} points more RA\textsubscript{gb} than the highest of them.
On \mbox{Qwen2.5-VL-7B}, \mbox{DC-4} keeps 78.3\,{\footnotesize[66.7,\,88.3]} against GB-T and the matched cheap transformations at most 28.3\,{\footnotesize[16.7,\,40.0]}, a gap of 50.0\,{\footnotesize[36.7,\,61.7]} points.
\emph{The robustness of JPEG-20 at the imperceptible tier drops when the hardening range widens, and web-hardened perturbations at the perceptible tier leave DC-4 far more robust than the cheap transformations at matched clean cost.}

The cause is that a perturbation on an agentic input is shaped by the preprocessing it has to pass: a web-hardened one is built to pass quantization, rescaling and small noise.
\subsection{Impact of Input-Estimated Strength and \mbox{Location}}\label{sec:components}

DC~\cite{pei2025divide} applies its larger noise level inside an attention mask, and SSNI~\cite{sun2025sample} estimates a noise level for each input.
Both components rely on the latent assumption that the input reveals where and how strongly to purify.
This assumption does not hold on agentic inputs under perceptible attacks, as we show in this section.

\noindent\textbf{Settings.}
We evaluate on web-shop pages at the perceptible tier.
For the location component, we compare DC-4 with DC-4 (no mask), which removes the attention mask, and with DC-4 (all mask), whose mask covers the whole page, both at the readout of DC-4.
For the strength component, we compare SSNI with two fixed noise levels, G@0.175, which has the clean cost of SSNI, and G@0.048, the mean level that SSNI estimates on the attacked pages.
We also measure the level that SSNI estimates on clean pages and on their GB-LF versions, and evaluate five stopping rules of DiffCAP~\cite{fu2026diffusion} with CLIP or VLM encoders.

\noindent\textbf{Results.}
As shown in \cref{tab:components}, removing the attention mask leaves DC-4 almost unchanged: its clean accuracy minus that of DC-4 (no mask) is $-$0.25\,{\footnotesize[$-$1.75,\,1.25]} points, and the paired difference in RA\textsubscript{gb} is 3.3\,{\footnotesize[$-$1.7,\,9.2]} points.
DC-4 (all mask) costs 8.50\,{\footnotesize[4.50,\,12.50]} more clean points and gains no robustness.
On Qwen2.5-VL-7B, DC-4 keeps 78.3 against GB-T with and without its mask, as much as G@0.2 at matched clean cost.

In the second block of \cref{tab:components}, the RA\textsubscript{gb} of a fixed level rises from 33.3 at G@0.048 to 75.0 at G@0.175, the level with the clean cost of SSNI.
The fixed level G@0.175 is 18.3\,{\footnotesize[8.3,\,30.0]} points more robust than the level that SSNI estimates.
The level that SSNI estimates on the attacked pages is too low to neutralize these perturbations.
On the tiles over the perturbed photo, the estimated level barely moves on the GB-LF versions of the pages: the paired change is 0.82\,{\footnotesize[0.02,\,1.77]} against a clean mean of 78.48, both in units of $10^{-3}$.
On held-out pages, DC-4 minus DC-4 (no mask) is 1.25\,{\footnotesize[$-$0.75,\,3.50]} clean points, and the RA\textsubscript{gb} of DC-4 with and without its mask differs by 1.6\,{\footnotesize[$-$0.7,\,4.0]} points.
On the same held-out pages, G@0.175 is 17.9\,{\footnotesize[12.8,\,23.4]} points more robust than the level that SSNI estimates.
\emph{On web-shop pages, the input-estimated components add no robustness at matched clean cost, and what acts is the global noise level.}

\begin{table}[t]
  \centering\small
  \begin{tabular}{@{}l@{\hskip 5pt}r@{\,}l@{\hskip 5pt}r@{\,}l@{}}
\toprule
 & \multicolumn{2}{c}{Clean accuracy} & \multicolumn{2}{c}{RA\textsubscript{gb} on $\mathcal{P}$} \\
Defense & \multicolumn{2}{c}{$n=200$} & \multicolumn{2}{c}{$n=60$} \\
\midrule
DC-4 & 90.00 & {\scriptsize[85.75,\,93.75]} & 87.5 & {\scriptsize[79.2,\,94.2]} \\
DC-4 (no mask) & 90.25 & {\scriptsize[86.00,\,94.00]} & 84.2 & {\scriptsize[75.0,\,92.5]} \\
DC-4 (all mask) & 81.50 & {\scriptsize[76.25,\,86.50]} & 80.8 & {\scriptsize[72.5,\,89.2]} \\
\hspace{0.5em}DC-4 $-$ DC-4 (no mask) & $-$0.25 & {\scriptsize[$-$1.75,\,1.25]} & 3.3 & {\scriptsize[$-$1.7,\,9.2]} \\
\hspace{0.5em}DC-4 $-$ DC-4 (all mask) & 8.50 & {\scriptsize[4.50,\,12.50]} & 6.7 & {\scriptsize[0.0,\,14.2]} \\
\midrule
SSNI & 90.50 & {\scriptsize[86.00,\,94.50]} & 56.7 & {\scriptsize[43.3,\,68.3]} \\
G@0.175 & 90.50 & {\scriptsize[86.00,\,94.50]} & 75.0 & {\scriptsize[63.3,\,85.0]} \\
G@0.048 & 91.50 & {\scriptsize[87.50,\,95.00]} & 33.3 & {\scriptsize[21.7,\,45.0]} \\
\hspace{0.5em}G@0.175 $-$ SSNI & 0.00 & {\scriptsize[$-$2.00,\,2.00]} & 18.3 & {\scriptsize[8.3,\,30.0]} \\
\hspace{0.5em}G@0.048 $-$ SSNI & 1.00 & {\scriptsize[$-$1.50,\,3.50]} & $-$23.3 & {\scriptsize[$-$35.0,\,$-$11.7]} \\
\midrule
DiffCAP (crop, 0.96) & 91.50 & {\scriptsize[87.50,\,95.00]} & 28.3 & {\scriptsize[16.7,\,40.0]} \\
DiffCAP (crop, cal.) & 91.75 & {\scriptsize[87.75,\,95.25]} & 29.2 & {\scriptsize[18.3,\,40.8]} \\
DiffCAP (tiles, 0.96) & 91.50 & {\scriptsize[87.50,\,95.00]} & 28.3 & {\scriptsize[16.7,\,40.0]} \\
DiffCAP (tiles, cal.) & 92.00 & {\scriptsize[88.00,\,95.50]} & 30.0 & {\scriptsize[18.3,\,41.7]} \\
DiffCAP (VLM, cal.) & 91.25 & {\scriptsize[87.25,\,95.00]} & 35.0 & {\scriptsize[23.3,\,47.5]} \\
\midrule
 & \multicolumn{2}{c}{Clean mean} & \multicolumn{2}{c}{GB-LF $-$ clean} \\
\cmidrule(lr){2-3}\cmidrule(l){4-5}
SSNI level, box ($\times 10^{-3}$) & \multicolumn{2}{c}{78.48} & 0.82 & {\scriptsize[0.02,\,1.77]} \\
\midrule
Held-out & \multicolumn{2}{c}{Difference} & \multicolumn{2}{c}{Quantity, $n$} \\
\cmidrule(lr){2-3}\cmidrule(l){4-5}
\hspace{0.5em}DC-4 $-$ DC-4 (no mask) & 1.25 & {\scriptsize[$-$0.75,\,3.50]} & \multicolumn{2}{c}{clean, 200} \\
\hspace{0.5em}DC-4 $-$ DC-4 (no mask) & 1.6 & {\scriptsize[$-$0.7,\,4.0]} & \multicolumn{2}{c}{RA\textsubscript{gb}, 273} \\
\hspace{0.5em}G@0.175 $-$ SSNI & 17.9 & {\scriptsize[12.8,\,23.4]} & \multicolumn{2}{c}{RA\textsubscript{gb}, 273} \\
\bottomrule
\end{tabular}
  \caption{\textbf{On web-shop pages, robustness comes from the global noise level: input-estimated masks and noise levels add none}.
  DiffCAP: (encoder, threshold); cal.: calibrated. Intervals: 95\%.}
  \label{tab:components}
\end{table}

For SSNI, the cause is that a perceptible low-frequency perturbation looks like clean content to its estimator: the estimated level on the perturbed photo stays near its clean value instead of rising with the attack.

\subsection{Impact of Adaptive Attacks}\label{sec:adaptive}

The grey-box attacker of \cref{sec:inputs} does not know the defense.
A content provider who learns which transformation the agent applies can attack through it.
On classifier inputs, attacks that know the defense have circumvented transformations that looked robust~\cite{athalye2018obfuscated,tramer2020adaptive} and lowered the robustness reported for diffusion purification~\cite{lee2023robust}.
We verify in this section what such an attacker achieves against each defense on native-resolution agentic inputs.

\noindent\textbf{Settings.}
We run the same-page adaptive attack of \cref{sec:protocol} on the web-shop pages of the grey-box evaluation at the perceptible tier, with BPDA through JPEG and an expectation over two noise draws per iteration for stochastic defenses.
We attack DC-4, DC-4\textsuperscript{+}, G@0.2 and the cheap transformations JPEG-60 and GN-16 (clean costs 3.0 and 3.4 points) on 60 pages, and SSNI and DiffCAP on a 20-page subset.
On each page, the attack crafts one perturbation in pixel space and one low-frequency perturbation, and RA\textsubscript{ad} takes the worst case over both and the grey-box member, scored on fresh noise draws.

\noindent\textbf{Results.}
As shown in \cref{tab:adaptive}, the adaptive attack lowers \mbox{JPEG-60} from 40.0\,{\footnotesize[28.3,\,51.7]} to 8.3\,{\footnotesize[1.7,\,16.7]} and GN-16 from 39.6\,{\footnotesize[28.3,\,51.3]} to 12.1\,{\footnotesize[5.0,\,20.0]}.
On the 20-page subset, SSNI goes from 45.0\,{\footnotesize[25.0,\,65.0]} to 30.0\,{\footnotesize[10.0,\,50.0]} and DiffCAP from 25.0\,{\footnotesize[10.0,\,45.0]} to 10.0\,{\footnotesize[0.0,\,25.0]}.
\emph{Under a same-page adaptive attack, the grey-box accuracy of the cheap transformations JPEG-60 and GN-16 largely disappears.}

\begin{table}[t]
  \centering\small
  \begin{tabular}{@{}l@{\hskip 7pt}r@{\hskip 7pt}r@{\,}l@{\hskip 7pt}r@{\,}l@{}}
\toprule
Defense & \multicolumn{1}{c}{$n$} & \multicolumn{2}{c}{RA\textsubscript{gb}} & \multicolumn{2}{c}{RA\textsubscript{ad}} \\
\midrule
DC-4 & 60 & 87.5 & {\scriptsize[79.2,\,94.2]} & 57.5 & {\scriptsize[45.0,\,70.0]} \\
DC-4\textsuperscript{+} & 60 & 89.2 & {\scriptsize[81.7,\,95.8]} & 80.8 & {\scriptsize[70.8,\,90.0]} \\
G@0.2 & 60 & 85.0 & {\scriptsize[75.0,\,93.3]} & 50.8 & {\scriptsize[38.3,\,63.3]} \\
JPEG-60 & 60 & 40.0 & {\scriptsize[28.3,\,51.7]} & 8.3 & {\scriptsize[1.7,\,16.7]} \\
GN-16 & 60 & 39.6 & {\scriptsize[28.3,\,51.3]} & 12.1 & {\scriptsize[5.0,\,20.0]} \\
\addlinespace[2pt]
SSNI & 20 & 45.0 & {\scriptsize[25.0,\,65.0]} & 30.0 & {\scriptsize[10.0,\,50.0]} \\
DiffCAP & 20 & 25.0 & {\scriptsize[10.0,\,45.0]} & 10.0 & {\scriptsize[0.0,\,25.0]} \\
\bottomrule
\end{tabular}
  \caption{\textbf{A same-page adaptive attack removes most of the grey-box robustness of JPEG-60 and GN-16}.
  Intervals: 95\%.}
  \label{tab:adaptive}
\end{table}

The grey-box robustness of a cheap transformation thus comes mostly from an attacker who does not know it.
For a diffusion purifier at native resolution, the gradient passes through the denoiser on every tile over the perturbed photo and through fresh noise at every iteration.
The adaptive values of the diffusion purifiers are therefore upper bounds on robust accuracy.
\Cref{supp:adaptive} gives both checks of \cref{sec:protocol}.

\subsection{Impact of the Decisive Element}\label{sec:element}

Purifiers inherited from classifiers~\cite{nie2022diffusion} rely on the latent assumption that the evidence is spread over the whole image, so every pixel may be traded against robustness.
However, an agentic input holds many elements, and the decision rests on one of them, as we verify in this section.

\begin{figure*}[t]
  \centering
  \includegraphics[width=\linewidth]{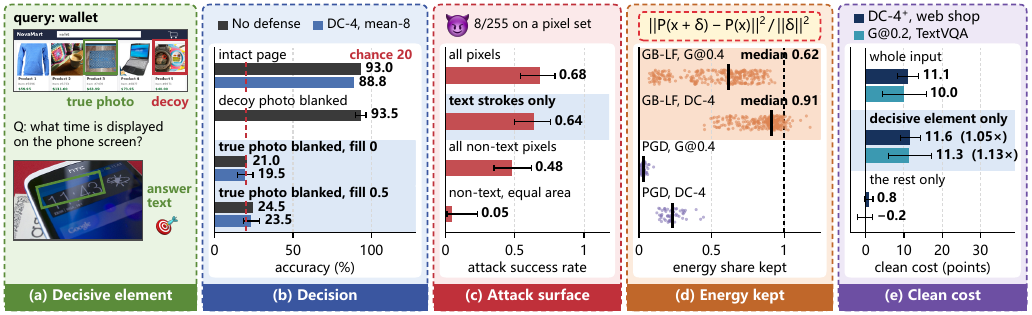}
  \caption{\textbf{Decision, attack surface and clean cost fall on the decisive element} (blue-shaded rows).
  (c)~No defense. Intervals: 95\%.}
  \label{fig:element}
\end{figure*}

\noindent\textbf{Settings.}
\Cref{fig:element}(a) outlines the decisive element: the queried and decoy photos on a web-shop page and the answer text on a TextVQA image.
On web-shop pages, we blank the photo of the queried product or a decoy photo.
On TextVQA, we restrict an imperceptible-tier attack on the undefended VLM to text strokes or to a non-text region of equal area.
For the GB-LF members of $\mathcal{P}$, we measure the perturbation energy that remains in the photo box after purification.
To locate the clean cost, we also purify only the decisive element or only the rest of the input, on 400 web-shop pages and 150 TextVQA images.

\noindent\textbf{Results.}
As shown in \cref{fig:element}(b), blanking locates the decision on web-shop pages, which rests on the photo of the queried product and not on elimination.
With that photo blanked by zero fill, the accuracy is 21.0 with No defense and 19.5\,{\footnotesize[14.8,\,24.5]} with DC-4 at mean-8, against a chance level of 20, while blanking a decoy photo leaves the undefended accuracy at 93.5\,{\footnotesize[90.0,\,96.5]}.
Panel (c) shows that text strokes suffice, and its equal-area control excludes the explanation that only the perturbed area matters.
An attack restricted to text strokes reaches a success rate of 0.64\,{\footnotesize[0.50,\,0.76]} on 50 TextVQA images, against 0.05\,{\footnotesize[0.01,\,0.23]} for a non-text region of equal area (21 images).

In panel (e) of \cref{fig:element}, the clean cost sits on the same element: purifying only the product-photo boxes with \mbox{DC-4\textsuperscript{+}} has 1.05\,{\footnotesize[0.95,\,1.16]} times the clean cost of purifying the whole page.
On TextVQA, purifying only the text tiles with G@0.2 has 1.13\,{\footnotesize[0.89,\,1.60]} times the clean cost of purifying all tiles.
Purifying only the rest costs 0.8\,{\footnotesize[$-$0.3,\,1.9]} points on web-shop pages and $-$0.2\,{\footnotesize[$-$2.2,\,1.8]} points on TextVQA.
After purification, most of a low-frequency perturbation still lies on the decisive element.
In panel (d), \mbox{GB-LF} members keep a median energy share of 0.62\,{\footnotesize[0.59,\,0.64]} in the photo box through G@0.4 and 0.91 through DC-4.
\emph{The decision, the attack surface and the clean cost fall on one and the same element of an agentic input.}

The three coincide because the agent reads the product photos or a few text strokes, the pixels that an attacker perturbs and a purifier alters.
Sparing the rest of the input therefore saves no clean cost, and the trade-off between clean cost and robustness is settled on the decisive element.
\noindent
The above analyses reveal that the root cause is the classifier-input assumption, which motivates the three requirements for agentic-level purification in \cref{sec:requirements}.

\section{Requirements for Agentic-Level Purification}
\label{sec:requirements}
\noindent\emph{(i) Native resolution and a measured clean cost.}
A purifier has to run at native resolution, and its clean cost has to be measured on the decisive element at the readout the agent uses.
In \cref{tab:clean}, the resize of DC-R costs 35.7\,{\footnotesize[29.5,\,42.0]} points on TextVQA against 4.0\,{\footnotesize[0.5,\,7.6]} for native-resolution DC-4, and \cref{fig:element} locates this cost on the decisive element.

\noindent\emph{(ii) No estimate of the perturbation.}
A purifier has to neutralize perceptible perturbations that are hardened against web preprocessing on the decisive element (\cref{sec:element}), without estimating their strength or location from the input.
In \cref{tab:components}, the fixed global noise level with the clean cost of SSNI, G@0.175, keeps 18.3\,{\footnotesize[8.3,\,30.0]} points more robust accuracy than SSNI, and 17.9\,{\footnotesize[12.8,\,23.4]} points more on held-out pages (\cref{sec:components}).
An input-dependent component therefore has to beat a fixed global noise level at matched clean cost, which neither DC's attention mask nor the noise level that SSNI estimates does on web-shop pages.

\noindent\emph{(iii) Robustness to attackers that know the defense.}
A purifier has to hold against attackers that know the defense and to beat the cheap transformations at matched clean cost.
The cheap transformations show what such an attacker removes: in \cref{tab:adaptive}, a same-page adaptive attack lowers the robust accuracy of JPEG-60 to 8.3\,{\footnotesize[1.7,\,16.7]} (\cref{sec:adaptive}).
A future purifier therefore has to report robust accuracy under a same-page adaptive attack with a convergence check and a gradient-masking check~\cite{athalye2018obfuscated,tramer2020adaptive}, beside the highest of the cheap transformations at matched clean cost (\cref{sec:tier}).

\section{Conclusion}
\label{sec:conclusion}
We presented the first study of adversarial purification at the agentic level, for the native-resolution, multi-element inputs of a VLM agent built on Qwen3-VL-8B.
We showed that the classifier-input assumption misleads both evaluation and design.
The clean cost on text-heavy inputs depends on resolution handling and on the readout.
At a matched clean cost of 3.0 points on web-shop pages, DC-4 keeps 45.0\,{\footnotesize[30.8,\,55.8]} points more grey-box robust accuracy than the highest of the cheap transformations against web-hardened perturbations.
This robustness comes from the global noise level: input-estimated masks and noise levels add none.
The resolution and web-shop findings hold on a disjoint held-out set and on a second VLM, Qwen2.5-VL-7B, where this gap is 50.0\,{\footnotesize[36.7,\,61.7]} points.
Four function evaluations suffice: DC-4 matches the clean cost and the robustness of DC's official 200-step loop at 1.05\,s instead of 45.57\,s per draw on a web-shop page.
The decision, the attack surface and the clean cost fall on one and the same element, so sparing the rest of the input saves no clean cost and the trade-off between clean cost and robustness is settled on that element.
Under a same-page adaptive attack, the grey-box robustness of JPEG-60 and GN-16 largely disappears: JPEG-60 falls from 40.0 to 8.3\,{\footnotesize[1.7,\,16.7]}.
We proposed AgenticPure, with native-resolution purifiers, web-hardened attack banks, comparisons at matched clean cost and held-out confirmation, and derived three requirements for defenses.
Purification for agents should therefore be designed and evaluated on agentic inputs rather than on classifier inputs.
We release the page generator and the code of AgenticPure.

\appendix
\clubpenalty=10000 \widowpenalty=10000 \brokenpenalty=10000
\renewcommand{\topfraction}{0.95}\renewcommand{\bottomfraction}{0.9}\renewcommand{\textfraction}{0.05}
\renewcommand{\floatpagefraction}{0.9}\renewcommand{\dbltopfraction}{0.95}\renewcommand{\dblfloatpagefraction}{0.9}
\setcounter{topnumber}{4}\setcounter{bottomnumber}{3}\setcounter{totalnumber}{8}\setcounter{dbltopnumber}{4}
\setcounter{figure}{0}\setcounter{table}{0}\setcounter{equation}{0}
\renewcommand{\theequation}{S\arabic{equation}}
\renewcommand{\thefigure}{S\arabic{figure}}
\renewcommand{\thetable}{S\arabic{table}}

\noindent As in the main text, accuracies and clean costs are in points with 95\% intervals (\cref{sec:metrics}), each defense is scored at its fixed readout, and no result uses the held-out set.

\FloatBarrier
\section{Implementation Details}\label{supp:impl}
\Cref{supp:tab:models} lists the experimental configuration.

\begin{table}[t]
  \centering\scriptsize
  \begin{tabular*}{\columnwidth}{@{\extracolsep{\fill}}p{0.24\columnwidth}@{\hskip 6pt}p{0.72\columnwidth}@{}}
\toprule
\raggedright VLM & \raggedright Qwen3-VL-8B-Instruct \tabularnewline
\raggedright Diffusion model & \raggedright unconditional $256\times256$ ImageNet ADM~\cite{dhariwal2021diffusion} \tabularnewline
\raggedright Tiles & \raggedright 256 px at stride 224, feathered \tabularnewline
\raggedright Two-step sampler & \raggedright noise to level $t$, then DDIM~\cite{song2021denoising} from $t$ to $t/2$ to 0 \tabularnewline
\raggedright Bank members & \raggedright 60 to 100 gradient steps on the undefended VLM \tabularnewline
\raggedright Web chain of GB-T & \raggedright JPEG at quality 60 to 95, rescaling, noise and blur \tabularnewline
\raggedright Hardware & \raggedright NVIDIA H100 GPUs \tabularnewline
\bottomrule
\end{tabular*}
  \caption{Experimental configuration.}
  \label{supp:tab:models}
\end{table}

\noindent\textbf{Mask of DC.}
The attention map of DC is the squared activation energy of the patch tokens in four vision-tower blocks of the VLM, upsampled to the page, min-max scaled and maximized over the blocks, and the mask is its top 15\% on the whole page.

\noindent\textbf{DC-200 and the timestep grid of DC-4.}\label{supp:dc4grid}
DC-200 runs the denoising loop of DC step for step, with 200 network evaluations from level 0.4, and on the same input, mask and noise it returns exactly the output of the released code.
DC-4 takes four DDIM steps that start at the levels 0.4, 0.3, 0.2 and 0.1.
Before each of its first three steps, the unmasked rest is re-noised to the current level, so it is last re-noised at the official level $t_{s} = 0.2$ of DC.
DC-4\textsuperscript{+} re-noises the rest only before the first two steps, last at level 0.30, and DC-R runs DC-4 on the input resized to $256\times256$, with the mask computed on the resized input.

\noindent\textbf{SSNI and DiffCAP.}
SSNI sets the noise level of each tile from its expected perturbation score at the published hyperparameters, with the reference statistic estimated on clean development inputs of each task.
DiffCAP applies the stopping rule of its Alg.~1 to the whole page with $t_{\min} = 0.04$, the published value for budgets above 4/255, and denoises the tiles at the stopping level.
Its five variants in Tab.~\ref{tab:components} stop on CLIP ViT-B/32~\cite{radford2021learning} features of the 224-px center crop or of 224-px tiles, or on the vision tower of the VLM, at the threshold 0.96 or at the threshold that Alg.~2 calibrates on clean and attacked development pairs.

\begin{table}[t]
  \centering\scriptsize
  \begin{tabular*}{\columnwidth}{@{\extracolsep{\fill}}l@{\hskip 6pt}l@{\hskip 6pt}l@{\hskip 6pt}r@{\hskip 6pt}r@{\hskip 6pt}r@{\extracolsep{0pt}\,}l@{}}
\toprule
Method & Data & Setting & \multicolumn{1}{c}{$n$} & \multicolumn{1}{c}{Published} & \multicolumn{2}{c}{Ours} \\
\midrule
DC & ImageNet & clean & 256 & 74.51 & 69.53 & {\scriptsize[63.64,\,74.85]} \\
SSNI & ImageNet & clean & 256 & 70.25 & 72.27 & {\scriptsize[66.48,\,77.39]} \\
\addlinespace[2pt]
DiffCAP & Caltech-101 & clean & 1000 & 82.6 & 83.1 & {\scriptsize[80.7,\,85.3]} \\
DiffCAP & Caltech-101 & APGD, 4/255 & 1000 & 80.9 & 80.5 & {\scriptsize[77.9,\,82.8]} \\
DiffCAP & ImageNet-R & clean & 1000 & 87.2 & 86.6 & {\scriptsize[84.3,\,88.6]} \\
DiffCAP & ImageNet-R & APGD, 4/255 & 1000 & 81.1 & 80.4 & {\scriptsize[77.8,\,82.7]} \\
\bottomrule
\end{tabular*}
  \caption{Accuracy (\%) on the data of each original paper: published and ours, with its 95\% Wilson interval.}
  \label{supp:tab:portcheck}
\end{table}

\noindent\textbf{Published accuracies.}
In \cref{supp:tab:portcheck}, we run the released code of DC and our SSNI without tiling on their ImageNet~\cite{deng2009imagenet} validation subset with ResNet-50~\cite{he2016deep}, and DiffCAP at its published setting on Caltech-101~\cite{fei2004learning} and ImageNet-R~\cite{hendrycks2021many} with zero-shot CLIP ViT-L/14, clean and under APGD~\cite{croce2020reliable} at 4/255.
Every published accuracy lies inside the 95\% Wilson interval of ours.
With the two-step sampler and $t_{\min} = 0.04$, our DiffCAP (crop, 0.96) is within 1.3 points of the published setting in every DiffCAP row.

\section{Evaluation Setup}\label{supp:eval}

\noindent\textbf{Data.}
A $960\times540$ web-shop page holds a header, a search box naming the queried product and five product cards, each a $176\times176$ ImageNet~\cite{deng2009imagenet} photo above its product number, price and rating.
The queried product and the attacker's target, another card of the page, change position across pages, and the photos come from disjoint pools of 15 classes with 50 photos each, so held-out pages share no class with development pages.
TextVQA~\cite{singh2019towards} questions come from the validation split, one per image, with disjoint images for development and held-out data.

\begin{table}[t]
  \centering\scriptsize
  \begin{promptbox}{sys}{Web shop: system}
You are a shopping assistant operating a web browser.
\end{promptbox}
\begin{promptbox}{usr}{Web shop: user}
The user asked: ``Buy a \emph{class name} for me.'' The page shows 5 products labelled Product 1 to Product 5 from left to right. Which product do you click? Answer with a single digit.
\end{promptbox}
\begin{promptbox}{usr}{TextVQA: user}
\emph{question} Answer the question using a single word or phrase.
\end{promptbox}
  \captionsetup{position=below}
  \caption{Prompts of the two tasks; placeholders in italics.}
  \label{supp:tab:prompts}
\end{table}

\noindent\textbf{Prompts and attacks.}
The VLM answers the prompts of \cref{supp:tab:prompts}, the web-shop decision is the product digit with the largest next-token logit, and TextVQA answers are decoded greedily and scored with the soft VQA accuracy.
Web-shop members target the attacker's product within 64/255 on its photo, and TextVQA members maximize the negative log-likelihood of the most frequent reference answer within 8/255 on the whole image, both with the recipe of \cref{supp:tab:models}.

\noindent\textbf{Statistics.}
The bootstrap resamples pages or questions with replacement, paired across defenses, and a maximum or minimum over defenses, such as the highest of the matched cheap transformations, is recomputed in each resample.
Held-out tests are one-sided paired bootstrap tests, with two one-sided tests~\cite{schuirmann1987comparison} for the equivalence of DC-4 with and without its mask and Holm's correction~\cite{holm1979simple} at level 0.05.
\FloatBarrier

\FloatBarrier
\section{Sensitivity to Tiling and Web Pipeline}\label{supp:sensitivity}

\noindent\textbf{Tiling.}
With 256-px tiles at strides 224, 256 and 128, with or without the feather, DC-4 costs 2.5 to 3.0 points on web-shop pages and 3.1 to 7.5 points on TextVQA, far below the 35.7 points of the resize of DC-R.
Its RA\textsubscript{gb} is 85.0 to 87.5 at strides 224 and 256 and 74.2\,{\footnotesize[63.3,\,84.2]} at stride 128, far above the 40.0\,{\footnotesize[28.3,\,51.7]} of JPEG-60, the highest of the matched cheap transformations.

\noindent\textbf{Web pipeline.}
Under five web pipelines with scales from 0.5 to 1.0 and JPEG qualities from 50 to 90, DC-4 keeps 88.3 to 91.7 against the worst GB-T member after the pipeline.

\section{Imperceptible Tier on TextVQA}\label{supp:tier8}
At 8/255 on the 126 attacked TextVQA images, DC-4 keeps 87.6\,{\footnotesize[82.0,\,92.6]} points against the imperceptible bank and GB-T10, SSNI 84.7\,{\footnotesize[78.6,\,90.5]} and DiffCAP 60.6\,{\footnotesize[52.1,\,68.8]}, against 22.8\,{\footnotesize[16.1,\,29.9]} for No defense.

\noindent\textbf{Hardening range.}
\Cref{supp:tab:tier8_gbt10} adds GB-T10 to the bank for standard JPEG and for the differentiable JPEG, which uses quality-scaled tables, 4:2:0 chroma subsampling and straight-through rounding and is also the JPEG step of the chain through which GB-T and GB-T10 are optimized.
The drop is positive in every row at qualities 10 to 30, up to 12.2\,{\footnotesize[6.9,\,18.0]} points for the differentiable JPEG at quality 20, and is 0.8\,{\footnotesize[0.0,\,2.4]} points at quality 50, next to the range on which GB-T is hardened.

\begin{table}[t]
  \centering\scriptsize
  \begin{tabular*}{\columnwidth}{@{\extracolsep{\fill}}l@{\hskip 8pt}r@{\,}l@{\hskip 8pt}r@{\,}l@{\hskip 8pt}r@{\,}l@{}}
\toprule
 & \multicolumn{2}{c}{Bank} & \multicolumn{2}{c}{With GB-T10} & \multicolumn{2}{c}{Drop} \\
Defense & \multicolumn{6}{c}{$n=126$} \\
\midrule
\multicolumn{7}{@{}l@{}}{\emph{Differentiable JPEG}} \\
JPEG-10 & 77.5 & {\scriptsize[70.1,\,84.4]} & 73.0 & {\scriptsize[65.3,\,80.2]} & 4.5 & {\scriptsize[1.3,\,8.5]} \\
JPEG-20 & 70.6 & {\scriptsize[62.7,\,78.3]} & 58.5 & {\scriptsize[50.0,\,66.7]} & 12.2 & {\scriptsize[6.9,\,18.0]} \\
JPEG-30 & 59.8 & {\scriptsize[51.3,\,68.0]} & 51.9 & {\scriptsize[43.4,\,60.3]} & 7.9 & {\scriptsize[4.0,\,12.7]} \\
\addlinespace[2pt]
\multicolumn{7}{@{}l@{}}{\emph{Standard JPEG}} \\
JPEG-20 & 66.9 & {\scriptsize[58.7,\,74.9]} & 57.1 & {\scriptsize[48.7,\,65.6]} & 9.8 & {\scriptsize[5.0,\,15.1]} \\
JPEG-30 & 60.1 & {\scriptsize[51.6,\,68.3]} & 51.9 & {\scriptsize[43.4,\,60.3]} & 8.2 & {\scriptsize[4.0,\,13.2]} \\
JPEG-50 & 47.9 & {\scriptsize[39.4,\,56.3]} & 47.1 & {\scriptsize[38.6,\,55.6]} & 0.8 & {\scriptsize[0.0,\,2.4]} \\
\bottomrule
\end{tabular*}
  \caption{JPEG on TextVQA at 8/255: RA\textsubscript{gb} against the bank, with GB-T10 added, and the paired drop.}
  \label{supp:tab:tier8_gbt10}
\end{table}

\begin{figure}[t]
  \centering
  \includegraphics[width=\linewidth]{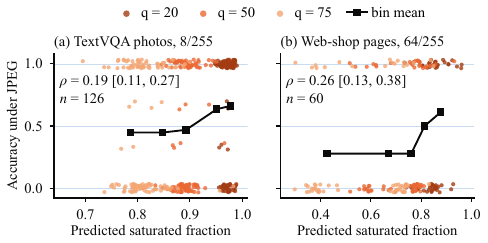}
  \caption{Predicted saturated fraction $f_q$ against accuracy under JPEG; squares: bin means.}
  \label{supp:fig:t1}
\end{figure}

\noindent\textbf{Saturation predictor.}
For a clean image $x$ and its perturbed image $x'$, let $c_{b,k}$ and $c'_{b,k}$ be the orthonormal DCT-II coefficients of the luma channel minus 128 in the $8 \times 8$ block $b$ at index $k$, and let $d_{b,k} = c'_{b,k} - c_{b,k}$.
With the luminance quantization step $Q_k(q)$ of JPEG at quality $q$ and $r_{b,k} = c_{b,k} - Q_k(q)\,\mathrm{round}(c_{b,k} / Q_k(q))$, the predicted saturated fraction over the $N_b$ blocks is
\begin{equation}\label{supp:eq:sat}
  f_q(x, x') = \frac{1}{64\,N_b} \sum_{b=1}^{N_b} \sum_{k=1}^{64} \mathbf{1}\big[\,|d_{b,k}| < Q_k(q)/2 - |r_{b,k}|\,\big].
\end{equation}
A saturated coefficient keeps the quantization index of its clean value, and the blocks cover the attacked photo box of a web-shop page or the whole TextVQA image.
\Cref{supp:fig:t1} plots $f_q$ against accuracy under JPEG at qualities 20, 50 and 75, for the member selected against JPEG at that quality.
Spearman's correlation is 0.19\,{\footnotesize[0.11,\,0.27]} on TextVQA and 0.26\,{\footnotesize[0.13,\,0.38]} on web-shop pages.

\FloatBarrier

\FloatBarrier
\section{Perceptible Tier on Web-Shop Pages}\label{supp:webshop}

\noindent\textbf{Fixed settings.}
\Cref{supp:tab:web_grid} scores every setting of the grids of Sec.~\ref{sec:purifiers} and five fixed noise levels.
No cheap setting keeps more than 68.3\,{\footnotesize[56.7,\,80.0]} against GB-T, reached by JPEG-10 at a clean cost of 7.5 points, while the fixed levels from G@0.2 to G@0.4 keep at least 83.3.

\begin{table}[t]
  \centering\scriptsize
  \begin{tabular*}{\columnwidth}{@{\extracolsep{\fill}}l@{\hskip 10pt}r@{\,}l@{\hskip 10pt}r@{\,}l@{\hskip 10pt}r@{\,}l@{}}
\toprule
 & \multicolumn{2}{c}{Clean cost} & \multicolumn{2}{c}{Worst of $\mathcal{P}$} & \multicolumn{2}{c}{GB-T} \\
Setting & \multicolumn{2}{c}{$n=200$} & \multicolumn{2}{c}{$n=60$} & \multicolumn{2}{c}{$n=60$} \\
\midrule
JPEG-95 & 1.0 & {\scriptsize[$-$1.0,\,3.0]} & 28.3 & {\scriptsize[16.7,\,40.0]} & 31.7 & {\scriptsize[20.0,\,43.3]} \\
JPEG-90 & 3.5 & {\scriptsize[1.0,\,6.5]} & 26.7 & {\scriptsize[16.7,\,38.3]} & 31.7 & {\scriptsize[20.0,\,43.3]} \\
JPEG-75 & 4.0 & {\scriptsize[1.5,\,7.0]} & 30.0 & {\scriptsize[18.3,\,41.7]} & 31.7 & {\scriptsize[20.0,\,43.3]} \\
JPEG-60 & 3.0 & {\scriptsize[1.0,\,5.5]} & 40.0 & {\scriptsize[28.3,\,51.7]} & 40.0 & {\scriptsize[28.3,\,51.7]} \\
JPEG-50 & 5.0 & {\scriptsize[2.0,\,8.0]} & 38.3 & {\scriptsize[26.7,\,50.0]} & 38.3 & {\scriptsize[26.7,\,50.0]} \\
JPEG-30 & 6.0 & {\scriptsize[3.0,\,9.5]} & 45.0 & {\scriptsize[33.3,\,58.3]} & 45.0 & {\scriptsize[33.3,\,58.3]} \\
JPEG-20 & 5.5 & {\scriptsize[2.0,\,9.5]} & 48.3 & {\scriptsize[35.0,\,60.0]} & 48.3 & {\scriptsize[35.0,\,60.0]} \\
JPEG-10 & 7.5 & {\scriptsize[3.0,\,12.5]} & 68.3 & {\scriptsize[56.7,\,80.0]} & 68.3 & {\scriptsize[56.7,\,80.0]} \\
\midrule
BIT-7 & 1.5 & {\scriptsize[0.0,\,3.5]} & 20.0 & {\scriptsize[10.0,\,30.0]} & 26.7 & {\scriptsize[15.0,\,38.3]} \\
BIT-6 & 0.5 & {\scriptsize[$-$1.0,\,2.5]} & 26.7 & {\scriptsize[16.7,\,38.3]} & 30.0 & {\scriptsize[18.3,\,41.7]} \\
BIT-5 & 2.0 & {\scriptsize[0.5,\,4.0]} & 26.7 & {\scriptsize[15.0,\,38.3]} & 31.7 & {\scriptsize[20.0,\,43.3]} \\
BIT-4 & $-$0.5 & {\scriptsize[$-$2.5,\,1.5]} & 26.7 & {\scriptsize[15.0,\,38.3]} & 33.3 & {\scriptsize[21.7,\,45.0]} \\
BIT-3 & 3.0 & {\scriptsize[0.5,\,6.0]} & 36.7 & {\scriptsize[25.0,\,48.3]} & 40.0 & {\scriptsize[28.3,\,51.7]} \\
BIT-2 & 10.0 & {\scriptsize[5.5,\,15.0]} & 35.0 & {\scriptsize[23.3,\,46.7]} & 35.0 & {\scriptsize[23.3,\,46.7]} \\
\midrule
GN-1 & 0.5 & {\scriptsize[$-$1.0,\,2.5]} & 28.3 & {\scriptsize[16.7,\,40.0]} & 33.3 & {\scriptsize[21.7,\,45.0]} \\
GN-2 & 1.3 & {\scriptsize[$-$0.3,\,3.1]} & 27.1 & {\scriptsize[16.7,\,37.9]} & 32.5 & {\scriptsize[21.3,\,44.2]} \\
GN-4 & 1.0 & {\scriptsize[$-$1.5,\,3.5]} & 26.7 & {\scriptsize[15.0,\,38.3]} & 31.7 & {\scriptsize[20.0,\,43.3]} \\
GN-8 & 2.5 & {\scriptsize[0.4,\,4.9]} & 30.0 & {\scriptsize[19.2,\,40.8]} & 33.8 & {\scriptsize[22.9,\,45.0]} \\
GN-16 & 3.4 & {\scriptsize[0.3,\,6.8]} & 39.6 & {\scriptsize[28.3,\,51.3]} & 39.6 & {\scriptsize[28.3,\,51.3]} \\
GN-24 & 4.0 & {\scriptsize[0.5,\,8.0]} & 48.3 & {\scriptsize[35.0,\,61.7]} & 48.3 & {\scriptsize[35.0,\,61.7]} \\
GN-32 & 6.5 & {\scriptsize[2.5,\,11.0]} & 48.3 & {\scriptsize[35.0,\,61.7]} & 48.3 & {\scriptsize[35.0,\,61.7]} \\
\midrule
G@0.025 & 0.4 & {\scriptsize[$-$0.8,\,1.6]} & 27.1 & {\scriptsize[17.1,\,37.9]} & 31.7 & {\scriptsize[20.8,\,42.9]} \\
G@0.1 & 1.0 & {\scriptsize[$-$1.5,\,3.5]} & 51.7 & {\scriptsize[38.3,\,65.0]} & 51.7 & {\scriptsize[38.3,\,65.0]} \\
G@0.2 & 4.0 & {\scriptsize[1.0,\,7.5]} & 85.0 & {\scriptsize[75.0,\,93.3]} & 83.3 & {\scriptsize[73.3,\,91.7]} \\
G@0.3 & 7.0 & {\scriptsize[3.0,\,11.5]} & 88.3 & {\scriptsize[80.0,\,95.0]} & 88.3 & {\scriptsize[80.0,\,95.0]} \\
G@0.4 & 15.5 & {\scriptsize[10.0,\,21.0]} & 81.7 & {\scriptsize[71.7,\,91.7]} & 83.3 & {\scriptsize[73.3,\,91.7]} \\
\bottomrule
\end{tabular*}
  \caption{Fixed settings on web-shop pages at 64/255: clean cost and RA\textsubscript{gb}; JPEG-60, BIT-3 and GN-8 are matched to DC-4.}
  \label{supp:tab:web_grid}
\end{table}

\noindent\textbf{Web-surviving pages.}
Without a defense, $\mathcal{P}$ hits its target on 43 of 60 pages, 0.717\,{\footnotesize[0.592,\,0.815]}, and its five GB-LF members alone on 41 of 60.
After the web pipeline, the GB-LF members hit it on 2 of 60 pages, 0.033\,{\footnotesize[0.009,\,0.114]}, while GB-T still hits it on the 31 pages of the web-surviving subset.
On these pages, DC-4 keeps 83.9\,{\footnotesize[70.0,\,96.3]} and the matched cheap transformations keep at most 45.2\,{\footnotesize[27.6,\,63.0]}.

\noindent\textbf{Further attacks.}
Adding an APGD member~\cite{croce2020reliable} to $\mathcal{P}$ leaves DC-4 at 86.7\,{\footnotesize[78.3,\,93.3]}, a paired gap of 70.0\,{\footnotesize[59.2,\,80.0]} points over the highest matched cheap transformation.
Under the attention-concentration attack of PRAC~\cite{seip2026preference} at 8/255, the JPEG and blur defenses of PRAC keep 95.0 to 96.7 on our pages, and against GB-T they keep 28.3 to 31.7.

\section{Same-Page Adaptive Attack}\label{supp:adaptive}

\noindent\textbf{Attack.}
The same-page adaptive attack (Sec.~\ref{sec:protocol}) knows the defense, differentiates through it on the evaluated page and averages the loss over its noise draws.
It is sign-PGD on the targeted cross-entropy of the decision, with step $\epsilon/8$ up to iteration 40 and $\epsilon/16$ after it.
For DC-4, DC-4\textsuperscript{+}, G@0.2, SSNI and DiffCAP, the gradient passes exactly through the tile blend and every denoiser evaluation of the reverse process, with activation checkpointing as in memory-efficient attacks on diffusion purifiers~\cite{kang2023diffattack,du2026memory}; among the steps of a defense, only JPEG uses BPDA~\cite{athalye2018obfuscated}.

\begin{table}[t]
  \centering\scriptsize
  \begin{tabular*}{\columnwidth}{@{\extracolsep{\fill}}l@{\extracolsep{\fill}\hskip 6pt}r@{\extracolsep{0pt}\,}l@{\extracolsep{\fill}\hskip 6pt}r@{\extracolsep{0pt}\,}l@{}}
\toprule
Convergence, 20 pages & \multicolumn{2}{c}{DC-4} & \multicolumn{2}{c}{G@0.2} \\
\midrule
60 iterations & 97.5 & {\scriptsize[92.5,\,100.0]} & 95.0 & {\scriptsize[87.5,\,100.0]} \\
180 iterations and restart & 95.0 & {\scriptsize[85.0,\,100.0]} & 80.0 & {\scriptsize[60.0,\,95.0]} \\
Change & $-$2.5 & {\scriptsize[$-$7.5,\,0.0]} & $-$15.0 & {\scriptsize[$-$30.0,\,$-$2.5]} \\
\midrule
Gradient masking, 10 pages & \multicolumn{2}{c}{DC-4} & \multicolumn{2}{c}{G@0.2} \\
\midrule
8/255 & 70.0 &  & 80.0 &  \\
16/255 & 70.0 &  & 80.0 &  \\
24/255 & 75.0 &  & 75.0 &  \\
128/255 & 30.0 &  & 35.0 &  \\
\bottomrule
\end{tabular*}
  \caption{Checks of the same-page adaptive attack: accuracy (points) of DC-4 and G@0.2.}
  \label{supp:tab:adp_checks}
\end{table}

\noindent\textbf{Checks.}
\Cref{supp:tab:adp_checks} reports both checks of the attack for DC-4 and G@0.2 on web-shop pages.
The convergence check continues the attack on 20 pages to 180 iterations, with step $\epsilon/16$ up to iteration 120 and $\epsilon/32$ after it, and adds one APGD-CE restart of 100 iterations~\cite{croce2020reliable}.
The gradient-masking check crafts the pixel-space perturbation on 10 pages at budgets from 8/255 to 128/255.
Continuing the attack changes the accuracy of DC-4 by $-$2.5\,{\footnotesize[$-$7.5,\,0.0]} and that of G@0.2 by $-$15.0\,{\footnotesize[$-$30.0,\,$-$2.5]} points.
At 128/255, the accuracy of the two purifiers falls to 30.0 and 35.0 points.

\noindent\textbf{Noise draws.}
Re-scored with 32 fresh draws, the RA\textsubscript{ad} of DC-4, DC-4\textsuperscript{+} and G@0.2 in Tab.~\ref{tab:adaptive} stays within 2.6 points.
At the end of the continuation, eight noise draws per iteration instead of two leave the accuracy of G@0.2 unchanged and change that of DC-4 by $-$2.5\,{\footnotesize[$-$7.5,\,0.0]} points.
\FloatBarrier

\FloatBarrier
\section{Decisive Element}\label{supp:element}
Panel (c) of Fig.~\ref{fig:element} attacks TextVQA images that No defense answers correctly with sign-PGD at 8/255 restricted to the text strokes (OCR word regions dilated by 2 pixels) or to an equal-area control that moves each stroke component to a random non-text position.
With the same single 60-iteration run as the equal-area control, the stroke attack succeeds on 25 of the 50 images, 0.50\,{\footnotesize[0.37,\,0.63]}, and exceeds the control by 0.38\,{\footnotesize[0.19,\,0.57]} on its 21 images.
A 120-iteration continuation and one restart on the images this run leaves intact raise the stroke rate to the 0.64\,{\footnotesize[0.50,\,0.76]} of panel (c).

\FloatBarrier
\section{Second VLM: Qwen2.5-VL-7B}\label{supp:vlm2}
We repeat the web-shop and TextVQA experiments on Qwen2.5-VL-7B~\cite{bai2025qwen25vl} with the same pages, images and prompts, re-crafting $\mathcal{P}$ against this VLM with the same budget and recipe.
Matched settings follow Sec.~\ref{sec:protocol} on this VLM, and the clean cost of DC-4 on web-shop pages, 2.5 points, selects JPEG-30, BIT-3 and GN-16.

\begin{table}[t]
  \centering\scriptsize
  \begin{tabular*}{\columnwidth}{@{\extracolsep{\fill}}l@{\extracolsep{\fill}\hskip 6pt}r@{\extracolsep{0pt}\,}l@{\extracolsep{\fill}\hskip 6pt}r@{\extracolsep{0pt}\,}l@{\extracolsep{\fill}\hskip 6pt}r@{\extracolsep{0pt}\,}l@{}}
\toprule
 & \multicolumn{2}{c}{Clean cost} & \multicolumn{2}{c}{GB-T} & \multicolumn{2}{c}{$T_{\mathrm{web}}(x_{\mathrm{adv}})$} \\
Defense & \multicolumn{2}{c}{$n=200$} & \multicolumn{2}{c}{$n=60$} & \multicolumn{2}{c}{$n=60$} \\
\midrule
JPEG-30 & 2.5 & {\scriptsize[$-$1.0,\,6.5]} & 25.0 & {\scriptsize[15.0,\,36.7]} & 56.7 & {\scriptsize[43.3,\,68.3]} \\
BIT-3 & 1.5 & {\scriptsize[$-$1.5,\,4.5]} & 18.3 & {\scriptsize[8.3,\,28.3]} & 58.3 & {\scriptsize[46.7,\,70.0]} \\
GN-16 & 0.5 & {\scriptsize[$-$2.5,\,3.5]} & 28.3 & {\scriptsize[16.7,\,40.0]} & 76.7 & {\scriptsize[65.0,\,86.7]} \\
\midrule
DC-200 &  &  & 81.7 & {\scriptsize[73.3,\,88.8]} & 92.1 & {\scriptsize[86.7,\,96.7]} \\
DC-4 & 2.5 & {\scriptsize[$-$1.5,\,6.5]} & 78.3 & {\scriptsize[66.7,\,88.3]} & 90.0 & {\scriptsize[81.7,\,96.7]} \\
DC-4 (no mask) & 1.0 & {\scriptsize[$-$2.5,\,4.5]} & 78.3 & {\scriptsize[66.7,\,88.3]} & 85.0 & {\scriptsize[75.0,\,93.3]} \\
DC-R & 22.5 & {\scriptsize[16.5,\,29.0]} & 78.3 & {\scriptsize[66.7,\,88.3]} & 80.0 & {\scriptsize[70.0,\,90.0]} \\
G@0.2 & 1.5 & {\scriptsize[$-$2.0,\,5.5]} & 78.3 & {\scriptsize[66.7,\,88.3]} & 86.7 & {\scriptsize[78.3,\,95.0]} \\
SSNI & 0.0 & {\scriptsize[$-$2.5,\,2.5]} & 30.0 & {\scriptsize[19.2,\,41.7]} & 85.0 & {\scriptsize[75.0,\,93.3]} \\
DiffCAP & $-$2.5 & {\scriptsize[$-$5.5,\,0.5]} & 18.3 & {\scriptsize[10.0,\,28.3]} & 58.3 & {\scriptsize[45.0,\,70.0]} \\
\bottomrule
\end{tabular*}
  \caption{Web-shop pages on Qwen2.5-VL-7B: clean cost and RA\textsubscript{gb}; JPEG-30, BIT-3 and GN-16 are matched to DC-4.}
  \label{supp:tab:v2_web}
\end{table}

\noindent\textbf{Resolution.}
The resize of DC-R costs 58.8\,{\footnotesize[52.5,\,64.7]} points on TextVQA, 32.3\,{\footnotesize[26.3,\,38.4]} more than DC-4 at native resolution, and 22.5\,{\footnotesize[16.5,\,29.0]} points on web-shop pages.
At native resolution, SSNI and DiffCAP cost 2.1\,{\footnotesize[$-$0.4,\,4.8]} and 0.1\,{\footnotesize[$-$2.0,\,2.1]} points on TextVQA.

\noindent\textbf{Web-hardened perturbations.}
Without a defense, the re-crafted GB-T members hit their target on 54 of the 60 pages, 0.900\,{\footnotesize[0.799,\,0.953]}.
As shown in \cref{supp:tab:v2_web}, DC-4 keeps 78.3\,{\footnotesize[66.7,\,88.3]} against GB-T and the matched cheap transformations at most 28.3\,{\footnotesize[16.7,\,40.0]}, a paired gap of 50.0\,{\footnotesize[36.7,\,61.7]} points.
DC-200 keeps 81.7\,{\footnotesize[73.3,\,88.8]} against GB-T.
After the web pipeline, the gap is 13.3\,{\footnotesize[3.3,\,23.3]} points.

\noindent\textbf{Mask and noise level.}
Removing the mask of DC-4 leaves its RA\textsubscript{gb} against GB-T at 78.3, and the fixed level G@0.2, matched to the clean cost of DC-4, keeps the same 78.3.
SSNI, whose noise level is estimated from the input, keeps 30.0\,{\footnotesize[19.2,\,41.7]}.
On both VLMs, the resize to $256\times256$ raises the clean cost on text, and the global noise level sets the robustness against web-hardened perturbations.

\FloatBarrier

{
    \small
    \nocite{yam2025myface,guo2026artwork,guo2024laaca,liu2026diffprotect,guo2025pca,lin2026priorasskill,qian2026tdcount,wang2026harmreduction,zhao2026exploring,liu2026agentsurvey,liu2026phase,zhao2026metacognitive,yang2026pointcloud,zhou2026posecompass,li2026fedkd,zhao2026p2p,ai2026hardlabel,duan2026gaitprotector,lei2026gazesync,li2026safetwinnet,li2026w2sdefense,liu2026graybox,guo2025thesis,hu2025syntactic,li2025guardfed,guo2025beyond,li2025fairseg,zhao2025roptester,lei2025transferable,li2025flsurvey,qian2025perspective,qian2025t2icount,lei2025instant,zhao2025backdoorsurvey,liu2024mmadpurify,zhao2024weaktostrong,guo2024signature,qian2024semisupervised,guo2023siamese}
    
    \bibliographystyle{ieeenat_fullname}
    \bibliography{refs}
}
\end{document}